\documentclass[12pt]{article}
\usepackage{latexsym}
\usepackage{amsfonts}
\usepackage[mathscr]{eucal}
\usepackage{amssymb}
\usepackage{graphicx}
\usepackage{caption2}
\usepackage{psfrag}

\input epsf

\title{Application of the Large--$N_c$ limit to a Chiral Lagrangian
  with Resonances\thanks{UAB-FT-535 Report}}

\author{Oscar Cat{\`a}\thanks{UAB $\&$ IFAE, Edifici Cn, 08193 Bellaterra
    (Barcelona), Spain}}
\date{}
\begin{document}

\maketitle

\begin{abstract}
It is shown that the implementation of the Large--$N_c$ approximation helps to get insight into the structure of, in principle, any QCD-like theory. As an example, we will compute the NLO corrections to $L_{10}$  in
the chiral limit with a Lagrangian with Resonances.  
\end{abstract}
\vskip 0.5cm
\noindent
{\bf 1. Introduction}

\noindent
The QCD Lagrangian is assumed to encode the whole description of the
strong interactions. At least, QCD has proven to be successful in
describing its perturbative regime, corresponding to the high
energy limit of the theory. However, a useful description of the strong interactions at
low energies is still needed. The problem is two-fold: first of all,
the strong coupling constant $\alpha_s$ blows up, thus excluding 
all perturbative techniques. Secondly, quark and gluon fields are not
the appropriate variables at low energies, where one observes hadrons,
and there is not even a clue as how to connect these two sets of
variables, or, in other words, understand what lies behind the
process of hadronisation.

The efforts devoted to these topics for the last 40 years or so have
led to remarkable progress. On the one hand, the low energy regime of QCD has been successfully approached by means
of Chiral Perturbation Theory. On the other hand, Lattice QCD and
Large--$N_c$ QCD are two firmly-established attempts to fill the gap
between low and high energies. 

In particular, the Large-$N_c$ limit is currently profusely used not only to establish
a link between String theory and the Standard Model, but also to
extract information from the Standard Model itself. Our aim in this paper is
to show how the Large--$N_c$ approximation can be used to provide a solid
framework to QCD-like theories, both by enabling the use of perturbation
theory consistently and by constraining the (often) too large number of
free parameters.
\vskip 0.2cm
\noindent
{\bf 2. Chiral Perturbation Theory}

\noindent
Chiral Perturbation Theory ($\chi$PT) is the Effective Field Theory (EFT) of QCD in the
sector of the light quarks ($u$,$d$,$s$). It describes the interactions of
the pseudoGoldstone bosons{\footnote{i.e., $\pi$, $k$ and $\eta$. To
    be generically referred to as pions thereafter.}} under the
spontaneously broken chiral symmetry
\begin{equation}
SU_L(3)\times SU_R(3)\rightarrow SU_V(3)
\end{equation}
as a power series in momenta and masses. The first two terms read {\cite{Gass1}}:
\begin{equation}
\mathcal{L}_{\chi}^{(2)}(U,DU)=\frac{f_\pi^2}{4}\langle
D_{\mu}U^{\dagger}D^{\mu}U+U^{\dagger}\chi+\chi^{\dagger}U\rangle
\end{equation}
\begin{eqnarray}
\mathcal{L}_{\chi}^{(4)}(U,DU) & = & L_1\langle
D_{\mu}{U}^{\dagger}D^{\mu}U\rangle^2+L_2\langle
D_{\mu}{U}^{\dagger}D_{\nu}U\rangle\langle
D^{\mu}{U}^{\dagger}D^{\nu}U\rangle+\nonumber\\ & + &
L_3\langle D_{\mu}{U}^{\dagger}D^{\mu}U
D_{\nu}{U}^{\dagger}D^{\nu}U\rangle+L_4\langle
D_{\mu}{U}^{\dagger}D^{\mu}U\rangle\langle{U}^{\dagger}\chi+{\chi}^{\dagger}U\rangle+\nonumber\\
& + & L_5\langle
D_{\mu}{U}^{\dagger}D^{\mu}U({U}^{\dagger}\chi+{\chi}^{\dagger}U)\rangle+L_6\langle{U}^{\dagger}\chi+{\chi}^{\dagger}U\rangle^2+\nonumber\\
& + &
L_7\langle{U}^{\dagger}\chi-{\chi}^{\dagger}U\rangle^2+L_8\langle{\chi}^{\dagger}U{\chi}^{\dagger}U+{U}^{\dagger}\chi{U}^{\dagger}\chi\rangle-\nonumber\\
& - & iL_9\langle
F_R^{\mu\nu}D_{\mu}UD_{\nu}{U}^{\dagger}+F_L^{\mu\nu}D_{\mu}{U}^{\dagger}D_{\nu}U\rangle+L_{10}\langle{U}^{\dagger}F_R^{\mu\nu}UF_{L\mu\nu}\rangle+\nonumber\\
  & + & H_1\langle
  F_{R\mu\nu}F_R^{\mu\nu}+F_{L\mu\nu}F_L^{\mu\nu}\rangle+H_2\langle{\chi}^{\dagger}\chi\rangle
\end{eqnarray}
$f_{\pi}$ being the pion decay constant in the chiral limit,
$f_{\pi}=87\pm 3.5\,$ MeV, and the $L_i$ low-energy couplings encoding
the information of the heavy degrees of freedom (hadrons lying above
the pions) compatible with their quantum numbers.

As with any EFT, the previous expansion is only
valid up to some threshold ($ \Lambda_{\chi PT}\sim 1\,$GeV). In going to
higher energies one has to resort to more general frameworks such as
the Lattice or Large--$N_c$. Still, not so ambitious an approach would
be to build up another EFT by inserting new dynamical degrees
of freedom (hadrons), thus taking over $\chi$PT at a threshold $\mu^*$
and covering a wider range of energies.
\vskip 0.2cm
\noindent
{\bf 3. Adding Resonances}

\noindent
The key ingredient is to notice that the $L_i$ couplings
entering the chiral Lagrangian are also
order parameters of Chiral Symmetry Breaking, i.e., they do not get
contributions from the continuum of $QCD$. Therefore, there is hope to
saturate them with a discrete number of resonances, the most favoured
ones being the lowest-lying hadrons. This is supported by the
following argument, taken from {\cite{pichn}}: consider a set of resonances, whose propagators are given generically by
\begin{equation}
\frac{1}{q^2-M_{\cal{R}}^2}
\end{equation}
At low energies, they can be expanded in powers of $q^2$, contributing
to $L_i$ like
\begin{equation}
L_i\sim \sum_{\cal{R}}\frac{F_{\cal{R}}^2}{M_{\cal{R}}^2}
\end{equation}
Thus, the natural choice would be to insert the lightest hadrons,
since they bear the biggest impact on $L_i$. One such proposal
is {\cite{Sui1}}:
\begin{eqnarray}
{\cal{L}}_{\cal{R}}(V,A,S,P)&=&{\sum_{R=V,A}}\left[-\frac{1}{2}\langle {\nabla}^\lambda
R_{\lambda\mu}{\nabla}_\nu
R^{\nu\mu}- \frac{1}{2} M_R^2
R_{\mu\nu}R^{\mu\nu}\rangle\,\right]-\nonumber\\
&-&\sum_{R=V,A}\left[\frac{1}{2}\partial^{\lambda}R_{1,\lambda\mu}\partial_{\nu}R_1^{\nu\mu}\,+\,\frac{M_{R_1}^2}{4}R_{1,\mu\nu}R_1^{\mu\nu}\right]\nonumber\\
&+&{\sum_{R=S,P}}\left[\frac{1}{2}\, \langle
\nabla^\mu R \nabla_\mu R - M_R^2 R^2\rangle +\, \frac{1}{2}\, \left(
\partial^\mu R_1 \partial_\mu R_1 - M_{R_1}^2 R_1^2\right)\right]\nonumber\\
&+&\frac{F_V}{2\sqrt{2}}\,\langle\,V_{\mu\nu}f_{+}^{\mu\nu}\rangle\,+\,i\frac{G_V}{\sqrt{2}}\langle\,
V_{\mu\nu}u^{\mu}u^{\nu}\rangle\ +
\frac{F_A}{2\sqrt{2}}\,\langle A_{\mu\nu}f_{-}^{\mu\nu}\rangle\nonumber\\
&+& c_d \,\langle S u_{\mu} u^\mu \rangle\,
+\,c_m\,\langle S\chi_+\rangle + \tilde{c}_d \, S_1\,
\langle u_{\mu} u^\mu \rangle\,+\tilde{c}_m\,S_1\langle \chi_+\rangle\nonumber\\
&+& id_m\langle P\chi_-\rangle+i\tilde{d}_m P_1\langle\chi_+\rangle
\end{eqnarray} 
to be added to the chiral Lagrangian, such that the enlarged chiral
theory looks like:
\begin{equation}
{\cal{L}}_{\cal{\chi R}}(V,A,S,P,U,DU)={\cal{L}}_{\chi}^{(2)}(U,DU)
+{\tilde{{\cal{L}}}}_{\chi}^{(4)}(U,DU)+{\cal{L}}_{\cal{R}}(V,A,S,P)
\end{equation}
One should be careful in distinguishing $L_i$ from
$\tilde{L}_i$: the former contain information about resonances lying
above the pions, whereas the latter have the information of the
resonances lying above the $V$, $A$, $S$ and $P$ multiplets{\footnote{Caveat: Eq.(6) is not intended to be a complete
    Lagrangian in any sense. It is the most general {\it{chirally-symmetric}}
    theory with {\it{lowest-lying}} hadronic multiplets and
    interaction terms {\it{linear}} in the resonance fields. But
    obviously, there is no compelling reason to do so. Indeed
    there are several other proposals for such an
    EFT.}}$^{,}${\footnote{$V$, $A$, $S$ and $P$ stand for Vector,
    Axial, Scalar and Pseudoscalar sectors.}}.
\vskip 0.2cm 
\noindent
{\bf 4. Integrating out Resonances}

\noindent
Once the Lagrangian ${\cal{L}}_{\cal{\chi R}}$ is chosen, the natural
step is to integrate out the resonances. We are left with the original chiral lagrangian,
the coefficients being functions of hadronic parameters (masses and couplings), to be related
to the $L_i$'s through a matching procedure. The results are as
follows {\cite{Sui1}}:
\begin{displaymath}
\begin{array}{lcccccccccccc}
L_1=&\frac{G_V^2}{8M_V^2}&+&
&-&\frac{c_d^2}{6M_S^2}&+&\frac{{\tilde{c}}_d^2}{2M_{S_1}^2}&+& &+& &\\
L_2=&\frac{G_V^2}{4M_V^2}&+& &+& &+& &+& &+& &\\
L_3=&-\frac{3G_V^2}{4M_V^2}&+& &+&\frac{c_d^2}{2M_S^2}&+& &+& &+& &
\end{array}
\end{displaymath}
\begin{equation}
\begin{array}{lcccccccccccc}
L_4=& &+& 
&-&\frac{c_dc_m}{3M_S^2}&+&\frac{\tilde{c}_d\tilde{c}_m}{M_{S_1}^2}&+&
&+& &\\
L_5=& &+& &+&\frac{c_dc_m}{M_S^2}&+& &+& &+& &\\
L_6=& &+& &-&\frac{c_m^2}{6M_S^2}&+&\frac{\tilde{c}_m^2}{2M_{S_1}^2}&+&
&+& &\\
L_7=& &+& &+&
&+& &+&\frac{d_m^2}{6M_P^2}&-&\frac{\tilde{d}_m^2}{2M_{{\eta}_1}^2}&\\
L_8=& &+& &+&\frac{c_m^2}{2M_S^2}&+& &-&\frac{d_m^2}{2M_P^2}&+& &\\
L_9=&\frac{F_VG_V}{2M_V^2}&+& &+& &+& &+& &+& &\\
L_{10}=&-\frac{F_V^2}{4M_V^2}&+&\frac{F_A^2}{4M_A^2}&+& &+& &+& &+& &
\end{array}
\end{equation}
\noindent\\
Upon comparison with experiment one gets{\footnote{Numerical values
    for masses and couplings are given in {\cite{Sui1}}.}}:
\begin{center}
\begin{tabular}{|c|c|c|c|c|c|c|c|c|}
\hline
$L_i(M_{\rho})$ & experimental values ($\times 10^{-3}$) & $V$ & $A$ & $S$ & $S_1$ & $\eta_1$ & total \\
\hline \hline
$L_1$ &     $0.7\pm0.3$ &   $0.6$ & $0$ & $-0.2$ & $0.2$ & $0$ & $0.6$       \\
\hline
$L_2$ &     $1.3\pm0.7$ &   $1.2$ & $0$ & $0$ & $0$ & $0$ & $1.2$      \\
\hline
$L_3$ &            $-4.4\pm2.5$ & $-3.6$ & $0$ & $0.6$ & $0$ & $0$ & $-3.0$         \\
\hline
$L_4$ &             $-0.3\pm0.5$ & $0$ & $0$ & $-0.5$ & $0.5$ & $0$ & $0.0$        \\
\hline
$L_5$ &        $1.4\pm 0.5$ & $0$ & $0$ & $1.4$ & $0$ & $0$ & $1.4$    \\
\hline
$L_6$ &         $-0.2\pm0.3$ & $0$ & $0$ & $-0.3$ & $0.3$ & $0$ & $0.0$       \\
\hline
$L_7$ &          $-0.4\pm0.15$ & $0$ & $0$ & $0$ & $0$ & $-0.3$ & $-0.3$       \\
\hline
$L_8$ &           $0.9\pm0.3$  & $0$ & $0$ & $0.9$ & $0$ & $0$ & $0.9$     \\
\hline
$L_9$ &            $6.9\pm 0.2$ & $6.9$ & $0$ & $0$ & $0$ & $0$ & $6.9$         \\
\hline
$L_{10}$ &          $-5.2\pm 0.3$ & $-10.0$ & $4.0$ & $0$ & $0$ & $0$ &   $-6.0$       \\
\hline
\end{tabular}
\end{center}
\noindent\\
which shows a remarkable agreement, strongly supporting resonance
saturation. Still, there is a flaw in the argument: the value $\mu^*=M_{\rho}$ is, though very suggestive, completely
arbitrary. The right way to proceed would be to determine $\mu^*$ from
${\cal{L}}_{\chi R}$ at one loop
level. However, contrary to ${\cal{L}}_{\chi}$, ${\cal{L}}_{\chi R}$
has no obvious expansion parameter. One way out is the following:
since ${\cal{L}}_{\chi R}$ is meant to be an EFT of QCD, we can
resort to Large--$N_c$ methods.
\vskip 0.2cm 
\noindent
{\bf 5. The Large--$N_c$ limit}{\footnote{The interested reader is
    referred to {\cite{Hooft}}.}}

\noindent
As mentioned above, the Large--$N_c$ limit of QCD is a powerful
non-perturbative tool to explore QCD and, in a broader sense, the
Standard Model. The crucial point is to realise that, surprisingly as
it is, a $SU(N_c)$ gauge theory of quarks and gluons is dynamically
simplified if one takes $N_c$ to infinity while keeping $\alpha_s N_c$
fixed. This last condition ensures that the theory is well-behaved
and, even further, that it admits a systematic expansion in powers
of $1/N_c$. Among the several additional features of the Large--$N_c$ limit we shall be interested only in a few ones:\\
\begin{itemize}
\item The dynamical
simplicity shows up in the fact that it is a theory of free,
stable and non-interacting mesons, where the interactions are
restricted to tree-level exchanges of physical mesons. This in turn
implies that the only singularities of Green functions are poles at
the masses of physical mesons.\\
\item The mass difference between singlets and octets is
    $1/N_c$-suppressed, so that in the $N_c\rightarrow \infty$ limit
    hadronic multiplets are nonets.
\end{itemize}
Despite the fact that even the leading term in the $1/N_c$
expansion entails an unsurmountable
calculational effort, those qualitative features highlighted above
provide, as we will see in a moment, a great deal of information.
\vskip 0.2cm
\noindent
{\bf 6. Large--$N_c$ applied to the model}

\noindent
Both the chiral Lagrangian ${\cal{L}}_{\chi}$ and the Resonance
Lagrangian ${\cal{L}}_{\cal{R}}$ can be seen in the light of
Large--$N_c${\footnote{The reader is referred to {\cite{LeutN}} and {\cite{Peris}} for details.}}. The first step is to know the $N_c$-scaling of the
parameters, which can be determined straightforwardly by a matching
procedure.
\begin{eqnarray}
L_{10}&\qquad\,&{\cal{O}}(N_c)\nonumber\\
f_{\pi},F_V,F_A,G_V,c_d,{\tilde{c}}_d&\qquad\,&{\cal{O}}(\sqrt{N_c})\nonumber\\
M_V,M_A,M_S,M_{S_1}&\qquad\,&{\cal{O}}(1)
\end{eqnarray}
The next step is to put singlets and octets together into
nonets. In the chiral limit it simply amounts to replace {\cite{Witt}}
\begin{equation}
U\rightarrow
U\,e^{-i{\frac{\sqrt{2}}{\sqrt{3}}}{\frac{\eta_1}{f_{\pi}}}}
\end{equation}
while the other terms in ${\cal{L}}_{\chi R}$ remain unaltered. This furnishes ${\cal{L}}_{\chi R}$ with a consistent Large--$N_c$
framework, as first shown in {\cite{Peris}}. As a by-product, the very particular
singularity structure of Large--$N_c$ Green functions provides a set
of constraining equations for the hadronic
parameters{\footnote{{\cite{Gol}} and references therein.}}  
\begin{equation}
F_V=2G_V=\sqrt{2}F_A=\sqrt{2}f_{\pi};\qquad M_A=\sqrt{2}M_V=4\pi f_{\pi}\left(\frac{2\sqrt{6}}{5}\right)^{1/2}
\end{equation}
which induce a {\it{prediction}} for the values of $L_i$. 
\begin{equation}
6L_1=3L_2=-\frac{8}{7}L_3=\frac{3}{4}L_9=-L_{10}=\frac{3}{8}\frac{f_{\pi}^2}{M_V^2}=\frac{15}{8\sqrt{6}}\frac{1}{16\pi^2}
\end{equation}
This is
remarkable, since it provides a testing ground for the accuracy of the
Large--$N_c$ approximation. Alternatively, Eq.($12$) can be read as the imprint of a (still) hidden dynamical
symmetry of the Large--$N_c$ limit. To make the point clearer, one can set out an analogy with
$SU(5)$ GUT. The parallelism is as follows:
\vskip 0.6cm
\begin{center}
\begin{tabular}{|c|c|}
\hline
\multicolumn{2}{|c|}{$\mathbf{SU(5)}$} \\
\hline
$\sin^2{\theta_W}=\frac{3}{8}$ &
$\alpha_{SU(3)}(M_{GUT})=\alpha_{SU(2)}(M_{GUT})=\alpha_{U(1)}(M_{GUT})$\\
\hline
\end{tabular}
\end{center}
\vspace*{0.3 cm}
\begin{center}
\begin{tabular}{|c|c|}
\hline
\multicolumn{2}{|c|}{\bf{Hypothetical Symmetry of Large--$N_c$}} \\
\hline
$L_{10}=-\frac{15}{8\sqrt{6}}\frac{1}{16\pi^2}$ & $6L_1=3L_2=-\frac{8}{7}L_3=\frac{3}{4}L_9=-L_{10}=\frac{3}{8}\frac{f_{\pi}^2}{M_V^2}=\frac{15}{8\sqrt{6}}\frac{1}{16\pi^2}$\\
\hline
\end{tabular}
\end{center}
\vspace*{0.2 cm}
Getting insight into this underlying symmetry would be of great interest
in trying to understand what lies behind the Large--$N_c$ approximation.
\vskip 0.2cm
\noindent
{\bf 7. An Example: $L_{10}$}

\noindent
In this section we will illustrate how theories equipped with the Large--$N_c$ approximation work with a particularly simple application: the
determination of $L_{10}$ with ${\cal{L}}_{\chi R}$ at one
loop level{\footnote{We will follow closely
    {\cite{article}} from here onwards. For further explanation the
    reader is referred to {\cite{article}} and references therein.}}$^{,}${\footnote{We shall adopt the $1/N_c$ expansion throughout as our underlying
power expansion.}}. Additionally, this simple example may help clarify
whether resonance saturation, as advocated in {\cite{Sui1}}, survives
beyond the leading order. The way to proceed is to integrate out
resonances in ${\cal{L}}_{\chi R}$ up to one loop. Equivalently, one can select an
appropriate Green function ({\it{i.e.}}, $L_{10}$ should appear without any mixing
with the remaining $L_i$) and compute it to one loop with both Lagrangians
at hand, ${\cal{L}}_{\chi}$ and ${\cal{L}}_{\chi R}$. For the present work, the latter method will be adopted. 

Let us start with the two-point function 
\begin{equation}
\Pi_{LR}^{\mu\nu}(q)\delta_{ab}=
 2i\int d^4 x\,e^{iq\cdot x}\langle 0\vert
\mbox{\rm T}\left(L^{\mu}_a(x)R^{\nu}_b(0)^{\dagger} \right)\vert
0\rangle\,,
\end{equation}
where 
\begin{equation}
  R_a^{\mu}\left(L_a^{\mu}\right)=
\bar{q}(x)\gamma^{\mu}\ \frac{\lambda_a}{\sqrt{2}}\
\frac{(1\pm\gamma_5)}{2}\ q(x)\, ,
\end{equation}
are the helicity currents of QCD. In the chiral limit, Lorentz
invariance implies
\begin{equation}
\Pi_{LR}^{\mu\nu}(Q^2)=(g^{\mu\nu}q^2-q^{\mu}q^{\nu})\Pi_{LR}(Q^2)
\end{equation}
Moreover, $\Pi_{LR}(Q^2)$, computed with $\chi$PT, admits the
following power expansion in $Q^2$
\begin{equation}
\Pi_{LR}(Q^2)=\frac{f_{\pi}}{Q^2}+4L_{10}+{\cal{O}}(Q^2)
\end{equation}
Therefore, a suitable Green function to determine $L_{10}$ is given by
\begin{equation}
\frac{1}{4}\frac{d}{dQ^2}(Q^2\Pi_{LR}(Q^2))_{Q^2=0}
\end{equation}
One can now compute ($17$) up to one loop with ${\cal{L}}_{\chi}$ and
${\cal{L}}_{\chi R}$, thereby getting
automatically the matching conditions. We will skip the calculations
and refer the reader to {\cite{article}} for further details. The
expressions one ends up with are
\begin{equation}
L_{10}=\frac{F_A^2}{4M_A^2}-\frac{F_V^2}{4M_V^2}+{\tilde{L}}_{10}=-\frac{1}{4}\left(\frac{15}{32\pi^2\sqrt{6}}\right)
+{\tilde{L}}_{10} 
\end{equation}
at tree level, which is in agreement with the last line of ($8$), and  
\begin{eqnarray}\label{result}
4\ L_{10}^r(\mu)&=&-\frac{15}{32\pi^2\sqrt{6}}\nonumber\\
&&-\frac{3}{2}\frac{F_A^2}{f_\pi^2}\frac{1}{(4\pi)^2}\left(\frac{1}{2}-
\log\frac{M_A^2}{\mu^2}\right)+
\frac{3}{2}\frac{F_V^2}{f_\pi^2}\frac{1}{(4\pi)^2}\left(\frac{1}{2}-
\log\frac{M_V^2}{\mu^2}\right)\nonumber\\
&&-\frac{5}{(4\pi)^2}\frac{G_V^2}{f_\pi^2}
\left(-\frac{17}{30}-\log\frac{M_V^2}{\mu^2}\right)\nonumber\\
&&+\frac{3}{2}\frac{1}{(4\pi)^2}\left(-\frac{1}{3}-\log\frac{M_A^2}{\mu^2}\right)
+\frac{3}{2}\frac{1}{(4\pi)^2}\left(-\frac{1}{3}-\log\frac{M_V^2}{\mu^2}\right)
\nonumber\\
&&
-\frac{4}{3}\left(\frac{{\tilde{c}}_d}{f_{\pi}}\right)^2\frac{1}{(4\pi)^2}
\left(\frac{1}{6}+\log\frac{M_{S_1}^2}{\mu^2}\right)-\frac{10}{9}
\left(\frac{c_d}{f_{\pi}}\right)^2\frac{1}{(4\pi)^2}
\left(\frac{1}{6}+\log\frac{M_S^2}{\mu^2}\right)\nonumber\\
&&+\frac{1}{2}\frac{1}{(4\pi)^2}
\left(1+\log\frac{M_S^2}{\mu^2}\right)-\nonumber\\
&&-\frac{4}{9}\left(\frac{c_d}{f_{\pi}}\right)^2
\frac{1}{(4\pi)^2}\left[\frac{1}{6}+\log\frac{M_S^2}{\mu^2}+ 2B + 2B^2 - (2 B^3 + 3 B^2) \log\frac{M_S^2}{M_{\eta_{1}}^2}\right]\nonumber\\
&&+ 4\ {\tilde{L}}_{10}^r(\mu) \ ,
\end{eqnarray}
up to one loop{\footnote{
$B=\frac{M_{\eta_{1}}^2}{(M_S^2-M_{\eta_{1}}^2)}$.}}. Notice that we
have kept the ${\tilde{L}}_{10}$ coupling all the way. Now we are in a position to assess whether
resonance saturation takes place or not. $L_{10}^r$ is plotted in
figure 1 (solid lines), together with the $L_{10}$ running (dashed
lines) and $L_{10}$ tree level value (dot-dashed lines). Since the
slope of $L_{10}^r$ does not follow that of runnning $L_{10}$, one
concludes that ${\tilde{L}}_{10}^r$ cannot be dropped (in fact, its
$\mu$-dependence has to supply the right slope for
$L_{10}^r$). However, there can still be a value for $\mu$ (say $\mu^*$) at
which resonance saturation is not spoiled (intersection of
solid and dashed lines). Yet this value lies at $\mu^*\sim 380$ MeV,
too low a value to be realistic{\footnote{Remember that this
    $\mu^*$ was to be interpreted as the threshold for
    ${\cal{L}}_{\chi R}$ to take over the chiral
    Lagrangian.}}.\\ Therefore, the Large--$N_c$ approximation has allowed us to
survey ${\cal{L}}_{\chi R}$ at one loop level, showing, through a
non-zero ${\tilde{L}}_{10}$, that for resonance saturation to be fulfilled one has to supply additional
structure ({\it{i.e.}}, new couplings and most likely new resonance
  fields) to our starting Lagrangian ${\cal{L}}_{\chi R}$. 

\begin{figure}
\renewcommand{\captionfont}{\small \it}
\renewcommand{\captionlabelfont}{\small \it}
\centering
\psfrag{m}{$\mu$}
\includegraphics[width=4.5in]{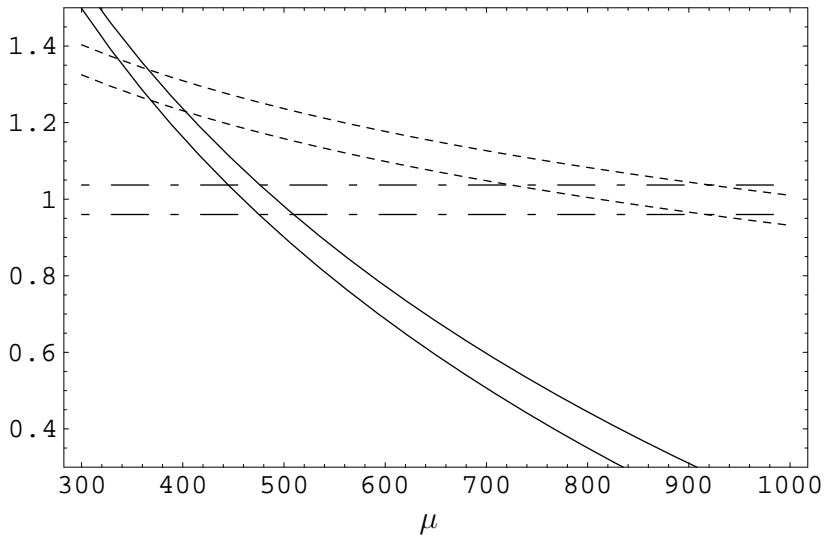}
\caption{$L_{10}(\mu)$ as a function of $\mu$. The solid lines
  represent $L_{10}^r(\mu)$ as given by Eq.(19) when imposing
  ${\tilde{L}}_{10}^r=0$. The dashed curves are the running of
  $L_{10}(\mu)$ as
  dictated by $\chi PT$. For comparison the tree level contribution to
  $L_{10}(\mu)$ is also given. All three curves are normalised to the
  central value of the tree level resonance contribution.}\label{fig:plot}
\end{figure}
\vskip 0.2cm
\noindent
{\bf 8. Conclusions}

\noindent
Taking Large--$N_c$ for granted, any QCD-like theory admits, in principle, a
perturbative series expansion in powers of $1/N_c$. This turns out to be
extremely useful when the theory hasn't got a natural perturbative
parameter, as in the example studied. At the same time, the
Large--$N_c$ limit is able
to constrain the values of the free parameters of the theory, these
constraints hinting at an underlying symmetry principle of Large--$N_c$.\\
\newline\\
\newline
\noindent
{\bf Acknowledgements}

\noindent
I would like to thank the organisers of the International School of
Subnuclear Physics (Erice '02), Professors A. Zichichi and G.'t Hooft, for giving me the
opportunity to deliver a talk.


\noindent


\begin{thebibliography}{99}


\bibitem{Gass1} J.~Gasser and H.~Leutwyler, Ann. of Phys. {\bf{158}}
    (1984) 142. J.~Gasser and H.~Leutwyler, Nucl. Phys. {\bf{B250}} (1985) 465.\bibitem{pichn} A.~Pich, [hep-ph/0205030]. To appear in the proceedings
         of The Phenomenology of Large $N(c)$ QCD, Tempe, Arizona,
         9-11 Jan 2002, Ed. R. Lebed, World Scientific, Singapore,
         2002.
\bibitem{Sui1}
G.~Ecker, J.~Gasser, A.~Pich and E.~de Rafael,
Nucl.\ Phys.\ {\bf{B321}} (1989) 311.
\bibitem{Hooft}
G.~'t Hooft,
Nucl.\ Phys.\ {\bf{B72}} (1974) 461.
E.~Witten,
Nucl.\ Phys.\ {\bf{B160}} (1979) 57.

\bibitem{LeutN} H. Leutwyler,
    Nucl.~Phys.~Proc.~Suppl.~{\bf{64}} (1998) 223
    [hep-ph/9709408]. R.~Kaiser and H.~Leutwyler,
    Eur.~Phys.~J.~{\bf{C17}} (2000) 623 [hep-ph/0007101].
\bibitem{Peris}
S.~Peris, M.~Perrottet and E.~de Rafael,
JHEP{\bf 9805} (1998) 011 [hep-ph/9805442].

\bibitem{Witt}
E.~Witten,
Nucl.\ Phys.\ {\bf{B156}} (1979) 269.
\bibitem{Gol} M.~F.~Golterman and S.~Peris,
Phys.\ Rev.\ D {\bf 61} (2000) 034018 [hep-ph/9908252].


\bibitem{article} O.~Cat{\`a} and S.~Peris, Phys.~Rev.~{\bf{D65}} (2002)
    056014 [hep-ph/0107062].



\end{thebibliography}
\end{document}